\documentclass[twocolumn]{revtex4-1}

\usepackage{amsfonts}
\usepackage{enumerate}
\usepackage{color}

\newtheorem{pro}{Problem}

\begin{document}
\title{The $k$-local Pauli Commuting Hamiltonians Problem is in P}
\author{Jijiang Yan}
\email{jjyan@cs.washington.edu}
\affiliation{Department of Physics, University of Washington, Seattle, WA 98195}
\affiliation{Department of Computer Science \& Engineering, University of Washington, Seattle, WA 98195}
\author{Dave Bacon}
\email{dabacon@google.com}
\altaffiliation[Current affiliation:]{ Google Inc.}
\affiliation{Department of Physics, University of Washington, Seattle, WA 98195}
\affiliation{Department of Computer Science \& Engineering, University of Washington, Seattle, WA 98195}
\begin{abstract}
Given a Hamiltonian that is a sum of commuting few-body terms, the commuting Hamiltonian problem is to determine if there exists a quantum state that is the simultaneous eigenstate of all of these terms that minimizes each term individually.  This problem is known to be in the complexity class quantum Merlin-Arthur, but is widely thought to not be complete for this class.  Here we show that a limited form of this problem when the individual terms are all made up of tensor products of Pauli matrices is efficiently solvable on a classical computer and thus in the complexity class P.  The problem can be thought of as the classical  XOR-SAT problem over a symplectic vector space. 
This class of problems includes instance Hamiltonians whose ground states possess topological entanglement, thus showing that such entanglement is not always a barrier for the more general problem. 
\end{abstract}
\maketitle

\section{Introduction}

Finding the energy of the ground and excited states of a many-body quantum system is a canonical problem in physics.  While there are many simple cases where one can exactly analytically solve for the energies, in general one often needs to resort to computer simulation to get even approximate answers to these problems.  Even then it is often the case that this problem is computationally intractable as a function of the size of many-body problem.  Recently, motivated in large part by the birth of quantum computing (whose original motivation was in part due to the intractable nature of simulating quantum systems~\cite{Feynman82}), a new understanding of the computational complexity of these quantum many body problems has begun to emerge.  A seminal result in this literature is that of Kitaev~\cite{KitaevTalk99,KSV02} who showed, roughly, that the problem of determining whether a system has a ground state energy below or above an energy level (specified to polynomial accuracy) is complete for a quantum equivalent of the computational complexity class  NP (non-deterministic polynomial time), the class QMA (quantum Merlin-Arthur).  This means that the problem is both as hard as every problem in QMA, and that if it could be solved efficiently it would allow for efficient algorithms for every problem in QMA.  Or, more loosely, that it is unlikely that even a quantum computer could help with this problem.  

Since Kitaev's pioneering work tremendous progress has been achieved in trying to understand the conditions under which this problem remains QMA-complete~\cite{AN02,BV04,Bravyi06,KKR06,ADKLLR07,Oliveira09,Schuch11,AE11}.  For example if the Hamiltonian of the many-body system is made up of a sum of commuting two-qubit or three-qubit interactions, then the problem is NP-complete instead of QMA-complete~\cite{BV04,AE11}.  The class of problems where the input Hamiltonians is made up of a sum of commuting terms has drawn special attention because these Hamiltonians can exhibit non-trivial entanglement (topological entanglement~\cite{Kitaev:06a}, for example) and yet according to physics lore should be thought of as essentially classical~\cite{BV04,AE11,Schuch11}.  Here we add another small result to this pantheon of results by showing that a quantum many-body system whose Hamiltonian is a sum of commuting terms that are made up of tensor products of Pauli matrices is solvable in polynomial time, i.e. is in the complexity class P.  While this is perhaps not surprising to those familiar with the theory of stabilizer error correcting codes~\cite{Gottesman97}, the Hamiltonians we consider include systems that exhibit topological order, such as Kitaev's toric code~\cite{Kitaev:03a}, thus perhaps shedding light on the role that topological order plays in the difficulty of the commuting Hamiltonian problem.  Our result can be thought of as instances of XOR-SAT over a symplectic vector space.

\section{The local Hamiltonian Problem}

Here we briefly review the local Hamiltonian problem.  We begin by defining the problem:
\begin{pro} {\em The $k$-local (qudit) Hamiltonian problem}
\noindent {\em Given:} A set of Hermitian positive  semi-definite operators, $\{H_i,1 \leq i \leq r\}$ acting on a Hilbert space of $n$ qudits (fixed dimension $d$) each of which is bounded norm, $\| H_i \| \leq c$, for some constant $c$.  Each term $H_i$ acts non-trivially on at most $k$ of the $n$ qudits and is specified to accuracy polynomial in $n$.  Additionally we are given two numbers, $b>a$ that are separated by an inverse polynomial gap, $b-a > 1/{\rm poly}(n)$.  We are promised that the smallest eigenvalue of $H=\sum_{i=1}^r H_i$ is either smaller than $a$ or greater than $b$.\\
\noindent {\em Decide:} Determine whether the smallest eigenvalue of $H=\sum_{i=1}^r H_i$ is smaller than $a$ or greater than $b$.
\end{pro}
The $k$-local Hamiltonian problem is meant, in part, to capture a natural class of problems that are encountered by physicists.  To this end the problem can be naturally thought of as attempting to find the energy of the ground state of a quantum-many body system.  Note however that the problem does not require actually producing an efficient description of the ground state.  

The $k$-local Hamiltonian was introduced by Kitaev~\cite{KitaevTalk99,KSV02} who showed that, with certain restrictions on the set of Hamiltonians, this problem is complete for the complexity class QMA.  Problems in QMA are, roughly, problems for which there exists a (quantum) proof such that if an instance is in the language then it can be efficiently verified on a quantum computer.  Thus Kitaev showed that for every ``yes'' instance of the $k$-local Hamiltonian problem there is an quantum state on {\rm poly}(n) qudits which can be used to verify that this instance is a ``yes'' instance.  Conversely Kitaev also showed that every problem that has an efficient quantum proof that can be verified in polynomial time on a quantum computer can be reduced into an instance of the $k$-local Hamiltonian problem.  

Kitaev's original result on the QMA-completeness of the $k$-local Hamiltonian problem placed certain restrictions on the Hamiltonian.  In particular his proof only held for qubit Hamiltonians with $k \geq 5$.  Subsequently this work was improved to $k=3$ (qubits)~\cite{KKR06} and $k=2$ (qubits)~\cite{Oliveira09}.  In another direction it was shown that the $k=2$ result holds even for Hamiltonians whose interactions graphs are planar~\cite{Oliveira09}, and even for quantum system in one spatial dimension when using qudits instead of qubits ($d=12$ in \cite{AGIK07}).  Thus a wide swath of Hamiltonians result in $k$-local Hamiltonian problems that are QMA-complete.

In a different direction one can also consider restrictions on the class of Hamiltonians being considered which reduce the complexity of the problem to (presumably) weaker complexity classes like NP and P.  For example if the Hamiltonian is diagonal in the computational basis of each qubit, then a seminal result of Barahona~\cite{Barahona:82a} shows that the problem is not QMA-complete, but instead is NP-complete.   Given this fact it is interesting to consider Hamiltonians that are made of sums of commuting $k$-local terms.  These can be simultaneously diagonalized so that in some basis they effectively behave like the classical problem, and yet because the diagonalization is non-trivial it is not clear that all of these problems are NP-complete.  This was investigated by Bravyi and Vyalyi~\cite{BV04} who showed that if the Hamiltonian is $2$-local then the problem is NP-complete.  Recently Aharonov and Eldar \cite{AE11} showed that this holds for $3$-local commuting Hamiltonians as well, for both qubits and qutrits.  This is especially nice because it shows that these systems do not possess topological order and in effect possess only local entanglement.  Motivated by this result in this paper we investigate a class of Hamiltonians that does possess topological order, those that are like Kitaev's toric code~\cite{Kitaev:03a} and made up of a sum of commuting Pauli operators.  In contrast to all of the problems above we here show that this problem is actually efficiently solvable in polynomial time.   

\section{The Commuting Hamiltonian Problem}

Having reviewed the $k$-local Hamiltonian problem, let us define the promise problem, $k$-local (qubit) Commuting Hamiltonian Problem:

\begin{pro} {\em $k$-local (qubit) Commuting Hamiltonian} \\
\noindent {\em Given:} A set of commuting $k$-local projectors on $n$ qubits, $\{H_j\}, 1 \leq j \leq r={\rm poly}(n)$ each of whose entries are specified using ${\rm poly}(n)$ bits of precision, and each of which pairwise commutes $[H_j,H_k]=0$ for all $j,k$.  We are promised that either
\begin{enumerate} [(a)]
\item There exists a zero energy eigenstate of $H=\sum_{j=1}^r H_j$, i.e. $\exists |\psi\rangle$ such that $H|\psi\rangle=0$.
\item There is no zero energy eigenstate and the lowest energy of eigenstates is at least $1$.  i.e. $\forall |\psi\rangle$ there exists a $j$ such that $\langle \psi |H_j|\psi\rangle \geq 1$ (this could be inverse polynomial, but in the case of commuting Hamiltonians the lowest energy, if it is not $0$, can be made into an equivalent problem with the lowest energy $\geq 1$.)
\end{enumerate}
{\em Decide:} Whether the given instance obeys condition (a) above.
\end{pro}

The commuting Hamiltonian problem is known to be NP-hard but is not known to be NP-complete.  A commonly used counter-example to the idea that commuting Hamiltonian problem is NP-complete are Hamiltonians such as Kitaev's toric code Hamiltonian.  These Hamiltonians are made up of commuting $k$-local terms yet possess a large amount of entanglement. In particular these Hamiltonians possess a global order, topological order, and hence cannot be prepared by a constant depth quantum circuit.  However, here we will show that this intuition is wrong and that the commuting Hamiltonian problem where all Hamiltonians that are made up of commuting projectors onto eigenspaces of $k$-qubit Pauli operators belongs, like the models considered by Kitaev, in fact, to the class P.  

Define the following problem, of which Hamiltonians like Kitaev's toric code are a particular example:
\begin{pro} {\em $k$-local Pauli Commuting Hamiltonian} \\
\noindent {\em Given:} A set of commuting $k$-local projectors on $n$ qubits, $\{H_j\}, 1 \leq j \leq r={\rm poly}(n)$ each of which pairwise commutes $[H_j,H_k]=0$ for all $j,k$  and for which each $H_j$ is a projector onto the $1$ eigenspace of a $k$-local Pauli operator, $H_i={1 \over 2}(I-S_i)$ where $S_i$ is a $k$-local Pauli operator with $S_i^2=I$.  We are promised that either
\begin{enumerate} [(a)]
\item There exists a zero energy eigenstate of $H=\sum_{j=1}^r H_j$, i.e. $\exists |\psi\rangle$ such that $H|\psi\rangle=0$.
\item There is no zero energy eigenstate and the lowest energy of eigenstates is at least $1$.  i.e. $\forall |\psi\rangle$, $\langle \psi |H|\psi\rangle \geq 1$
\end{enumerate}
{\em Decide:} Whether the given instance obeys condition (a) above.

\noindent  {\bf Claim:} $k$-local Pauli Commuting Hamiltonian problem is in P.
\end{pro}
Recall that the single qubit Pauli operators are $I=\left[\begin{array}{cc} 1 & 0 \\ 0 & 1 \end{array} \right]$, $X=\left[\begin{array}{cc} 0 & 1 \\ 1 & 0 \end{array} \right]$, $Y=\left[\begin{array}{cc} 0 &  -i\\ i & 0 \end{array} \right]$, and $Z=\left[\begin{array}{cc} 1 & 0 \\ 0 & -1 \end{array} \right]$ and that  Pauli operator on $n$ qubits is a tensor product of these operators (with a possible phase a multiple of $i$).

The proof of this claim will now occupy the rest of this section.  The main intuition is that this problem is related to a question about stabilizer codes, a question for which a sufficient Clifford circuit can transform this problem into an XOR-SAT problem, which is efficiently solvable on a classical computer.  Further this Clifford circuit can be efficiently computed and simulated.  Thus the essential tool is the Gottesman-Knill theorem~\cite{Gottesman99}.

As a first step we recast this problem in a slightly more palpable manner for those who know stabilizer codes.  In particular if we write $H_i={1 \over 2}(I-S_i)$, where $S_i$ is a $k$-local Pauli operator that squares to identity $S_i^2=I$, then the two conditions become
\begin{enumerate} [(a)]
\item $\exists |\psi\rangle$ such that $\forall i$, $S_i |\psi\rangle= |\psi\rangle$.
\item $\forall |\psi\rangle$ that are common eigenstates of the $S_i$'s, there exists at least one $i$ such that $ S_i |\psi \rangle = -|\psi\rangle$
\end{enumerate}
The problem now is given $S_i$ that are $k$-local Pauli operators, decide whether there is a common $+1$ eigenstate of these operators.  Here we will show how this can be done in polynomial time on a classical computer.  To be precise, define the 
\begin{pro} {\em $k$-local Pauli Commuting Hamiltonian (Stabilizer Version)} \\
\noindent {\em Given:} A set of commuting $k$-local Pauli operators, $\{S_j\}$, $1 \leq j \leq r={\rm poly}(n)$ which all square to identity.  We are promised that (actually it is always true that)
\begin{enumerate} [(a)]
\item $\exists |\psi\rangle$ such that $\forall i$, $S_i |\psi\rangle= |\psi\rangle$.
\item $\forall |\psi\rangle$ that are common eigenstates of the $S_i$'s, there exists at least one $i$ such that $ S_i |\psi \rangle = -|\psi\rangle$
\end{enumerate}
{\em Decide:} Whether the given instance obeys condition (a) above
\end{pro}
At this point we should note that we can further drop the condition of $k$-locality as so we will drop this from here on out.

We begin by reviewing some basic facts. Every Pauli operator that squares to identity can be written as $(-1)^p\otimes_{i=1}^n P(x_i,z_i)$ where, $p \in \{0,1\}$ and
\begin{equation}
P(x_i,z_i)=\left \{ \begin{array}{ll} I & {\rm if}~x_i=z_i=0 \\
X & {\rm if}~x_i=1,z_i=0\\
Y & {\rm if}~x_i=z_i=1\\
Z & {\rm if}~x_i=0,z_i=1
\end{array} \right.
\end{equation}
Thus we can associate to every such operator a $2n+1$ vector of binary numbers $(p,x_1,\dots,x_n,z_1,\dots,z_n)$.  For our $r$ $S_i$ operators we can denote these operators using $r$ $2n+1$ binary vectors, which we will denote $(p_i,x_{i,1},\dots,x_{i,n},z_{i,1},\dots,z_{i,n})$.  We can assemble these into a $r$ by $2n+1$ matrix, which we call $M$.

Now notice the following stability of our problem.  Suppose ${\mathcal S}=\{S_j|1 \leq j \leq r\}$ is an instance of our problem.  Let ${\mathcal S}^\prime$ denote the set formed by replacing the $k$th element of ${\mathcal S}$ by $S_j S_k$ where $j \neq k$.  We claim that ${\mathcal S}$ satisfy (a) iff ${\mathcal S}^\prime$ satisfy (a), and similarly for (b).  To see this note that if ${\mathcal S}$ satisfy (a), then $\exists |\psi\rangle$ such that $\forall i$ $S_i|\psi\rangle = |\psi \rangle$, and hence for this same $|\psi \rangle$, $\forall i \neq k, S_i|\psi\rangle = |\psi \rangle$, while $S_j S_k |\psi\rangle= |\psi\rangle$, since we can apply each $S_j$, $S_k$ in turn.  Similarly if ${\mathcal S}^\prime$ satisfy (a), then $\exists |\psi\rangle$ such that $S_i|\psi\rangle =|\psi\rangle$, $i \neq k$, and $S_j S_k |\psi\rangle=|\psi\rangle$, and hence for this same $|\psi\rangle$ $S_i|\psi\rangle=|\psi\rangle$ for all $i$, with the notable case that $S_j (S_j S_k) |\psi\rangle = S_k |\psi\rangle =|\psi\rangle$. 

For the no instances we proceed similarly. For convenience let's denote ${\mathcal S}^\prime$ by ${\mathcal S}^\prime = \{S_i|1 \leq i \leq r \mbox{ and } i \ne k\} \cup \{S_jS_k\}$, where $j \ne k$. If ${\mathcal S}$ satisfies (b), then $\forall |\psi\rangle$ that are common eigenstates of $S_i$'s, there exists at least one $S_i \in {\mathcal S}$ such that $S_i |\psi \rangle = -|\psi\rangle$. If $i \ne k$, then we are done, since there is at least this same $S_i$ in ${\mathcal S}^\prime$ such that $S_i |\psi \rangle = -|\psi\rangle$. If $i = k$, that is $S_k |\psi \rangle = -|\psi\rangle$, then there are two possibilities. One possibility is that $S_k$ is the only one in ${\mathcal S}$ such that $S_k |\psi \rangle = -|\psi\rangle$ and implicitly for all other $S_j \in {\mathcal S}$, $S_j |\psi \rangle = |\psi\rangle$. Then we would have $S_j S_k \in {\mathcal S}^\prime$ such that $S_j S_k |\psi \rangle = -S_j |\psi\rangle = -|\psi\rangle$. The other possibility is that there are other $S_j$'s such that $S_j |\psi \rangle = -|\psi\rangle$. Then we have at least those $S_j$'s in ${\mathcal S}^\prime$, no matter whether $S_j S_k |\psi \rangle = -|\psi\rangle$ or $S_j S_k |\psi \rangle = |\psi\rangle$. Hence if ${\mathcal S}$ satisfies (b), then ${\mathcal S}^\prime$ satisfies (b). In the other direction, if ${\mathcal S}^\prime$ satisfies (b), there are also two possibilities. If there is any $S_i \in {\mathcal S}^\prime$, where $i \ne k$, such that $S_i |\psi \rangle = -|\psi\rangle$ then we are done. Otherwise if $S_j S_k$ is the only one in ${\mathcal S}^\prime$ such that $S_j S_k |\psi \rangle = -|\psi\rangle$ and implicitly for all other $S_j \in {\mathcal S}^\prime$, $S_j |\psi \rangle = |\psi\rangle$. Then we have $S_k |\psi \rangle = -|\psi\rangle$ in ${\mathcal S}$. Hence we have shown that if ${\mathcal S}^\prime$ satisfies (b), then ${\mathcal S}$ satisfies (b).

Thus we have shown that an instance ${\mathcal S}$ can be converted into a completely equivalent instance, ${\mathcal S}^\prime$ by replacing an element of ${\mathcal S}$ with a product of that element and another element from ${\mathcal S}$.  Suppose that $M$ is the matrix describing the ${\mathcal S}$ as above and $M^\prime$ is the matrix describing ${\mathcal S}^{\prime}$.  Ignoring, for the moment, the first column of $M$ (corresponding to the phase factors), $M^{\prime}$ is simply the matrix for $M$ with the $k$th row replaced by the bit-wise addition modulo $2$ of row $j$ and row $k$ of $M$.  The phase factor can be calculated as follows.  First note that only the phase factor for the $k$th row of $M^\prime$ is (potentially) different from that of $M$.  Define
\begin{equation}
g(x_1,z_1,x_2,z_2)= \left \{  \begin{array}{ll} 0 & {\rm if}~x_1=z_1=0 \\
z_2-x_2 &{\rm if}~x_1=z_1=1 \\
z_2 (2x_2-1) & {\rm if}~x_1=1,z_1=0 \\
x_2 (1-2z_2) & {\rm if}~x_1=0,z_1=1 
\end{array} \right.
\end{equation}
Then if $2 r_j + 2 r_k + \sum_{i=1}^n g(x_{j,i},z_{j,i}, x_{k,i}, z_{k,i})=0~{\rm mod}~4$ set the first bit of the $k$th row of $M^\prime$ to $0$ otherwise set the first bit to $1$ (the sum will only ever give $2~{\rm mod}~4$.)  While this phase factor is a bit messy the only real important factor is that in going from ${\mathcal S}$ to ${\mathcal S}^\prime$ we can efficiently calculate the new phase factor of $M^\prime$ from $M$.  

Next note that our problem is invariant under unitary conjugation of the $\{S_j\}$ for a particular instance.  That is if we let ${\mathcal S}=\{S_j | 1 \leq j \leq r\}$, and ${\mathcal S}^\prime=\{U S_j U^\dagger, 1 \leq j \leq r\}$, then ${\mathcal S}$ satisfies (a) iff ${\mathcal S}^\prime$ satisfies (a) and similarly for (b).  This follows simply from the fact that unitary conjugation corresponds to a change of basis and both statements of (a) and (b) are basis-agnostic.  Further if the $U$ used is a Clifford group unitary, then the new set of operators ${\mathcal S}^\prime$ will all be elements of the Pauli group and thus in this case we obtain an equivalent Pauli Commuting Hamiltonians problem.  Suppose that $v$ is a $2n+1$ dimensional binary vector corresponding to a Pauli group element.  Then all elements of the Clifford group can be written as a linear transform over ${\mathbb Z}_2$ that acts on the last $2n$ elements of $v$ along with a transformation on the first bit (the global) phase that is a function of the elements of $v$ (this transform is not linear, but is easy to calculate.)  Not all linear transforms can be implemented this fashion since the structure of the Pauli group must be preserved.  

We will use in particular three transforms, the controlled-not $C_X$, the Hadamard $H$, and a square root of $Z$ gate $S={1 \over \sqrt{2}}(I-iZ)$.  The Hadamard on qubit $i$ simply has the effect of swapping the $X$ and $Z$ portions of the vector corresponding to the $i$th qubit, and changes the global phase bit $r$ to $r^\prime = r\oplus x_i z_i$ where $x_i$ is the $X$ component of the $i$th qubit and $z_i$ is the $Z$ component of the $i$th qubit. The $C_X$ gate, acting from the $i$th qubit to the $j$th qubit has the effect of replacing the $j$th qubit's $X$ component of the vector to the sum of this component and the $X$ component of the $i$th qubit, or $x_j \to x_i \oplus x_j$, while replacing the $i$th qubit's $Z$ component of the vector to the sum of this component and the $Z$ component of the $j$th qubit, or $z_i \to z_i \oplus z_j$.  The global phase is updated as a function of these values in a simple fashion, it is the sum of the old global phase plus the product of the $X$ component of the $i$th qubit $x_i$ times the $Z$ component of the $j$th qubit $z_j$ and the $X$ component of the $j$th qubit $x_j$ plus the $Z$ component of the $i$th qubit $z_i$ plus one,  or $r^\prime = r\oplus x_i z_j(x_j \oplus z_i \oplus 1)$.  Anyway it is easy to calculate function of the involved vector components.  The $C_X$, ignoring the global phase, thus has the effect of performing two simultaneous row additions, which row being added to which depends on which qubit is the target and which is the control.  The final gate, a square root of $Z$ gate, transforms Pauli operators as $SXS^\dagger = Y$, $SYS^\dagger = -X$, and $SZS^\dagger = Z$. Thus if we apply this gate to the $i$th qubit it has the effect of changing the $Z$ component of the $i$th qubit into the addition of the $Z$ component of the $i$th qubit and the $X$ component of the $i$th qubit, or $z_i \to x_i \oplus z_i$.  The global phase bit $r$ is updated to $r^\prime = r \oplus x_i z_i$. 

Now we are ready to describe the algorithm for solving the Pauli Commuting Hamiltonian problem, specified by a given $M$ matrix.  The tools that we have available to transform this $M$ to an equivalent problem are the ability to add rows, to perform the simultaneous column additions corresponding to the $C_X$ operator, the ability to swap the $X$ and $Z$ components, to permute both the $X$ and $Z$ components simultaneously using a permutation gate, and finally the ability to permute the rows of the matrix.  All of these can be done while efficiently updating the global phase bits for the matrix.  We will proceed by showing that we can use these operations to put $M$ into a form from which, either we have discovered along the way that the instance cannot satisfy (a) or there exists a $|\psi\rangle$ satisfying (a).

Let us denote the $M$ matrix by
\begin{equation}
M=\left[ \begin{array}{c||c||c} R & A & B \end{array} \right]
\end{equation}
where $R$ is the $r$ by $1$ matrix corresponding to the global phase bits, and $A$ and $B$ are the $r$ by $n$ matrices corresponding to the $X$ and $Z$ components of these operators respectively.  We now proceed as follows: by using permutations, $C_X$ and row sum, each of which preserves whether an instance satisfies (a) or not, we can perform Gaussian elimination on the matrix $A$.  When we do this, we will end up with a matrix of the form
\begin{equation}
M_1=\left[ \begin{array}{c||cc||cc} R_1 & I & 0  & B_1 & C_1  \\
R_2 & 0 & 0 & B_2 & C_2 \end{array} \right]
\end{equation}
where $I$ is a $k$ by $k$ identity matrix, with $k$ equal to the rank of $A$.  If, during this elimination, we ever obtain a row that has a global phase bit of $1$ and all zeros in the remaining rows, then we can immediately answer that the instance satisfies (b), because this corresponds to having the element $-I$ in the set of commuting Pauli operators for this instance, and it is trivial that (a) can never be satisfied in this case.  Using the fact that the operators corresponding to the rows must all commute we can further see that it must be the case that $B_2=0$:
\begin{equation}
M_1=\left[ \begin{array}{c||cc||cc} R_1 & I & 0  & B_1 & C_1  \\
R_2 & 0 & 0 & 0 & C_2 \end{array} \right]
\end{equation}
Furthermore, for the same reason, $B_1$ must be a symmetric matrix which could be further transformed into the form  $NN^T$, where $N$ is an invertible matrix~\cite{AG04},
\begin{equation}
M_1=\left[ \begin{array}{c||cc||cc} R_1 & I & 0  & NN^T & C_1  \\
R_2 & 0 & 0 & 0 & C_2 \end{array} \right]
\end{equation}
If we apply $C_X$ gate according to the form of $N$, we would obtain the new matrix 
\begin{equation}
M_1=\left[ \begin{array}{c||cc||cc} R_1 & N & 0  & N & C_1  \\
R_2 & 0 & 0 & 0 & C_2 \end{array} \right]
\end{equation}

Now if we apply $S$ gates on the first $k$ qubits, then the second $N$ matrix on the first block row would become zero. And if we perform Gaussian elimination again on the first $N$ matrix, we would transform it into $I$ and get matrix $M_2$, 
\begin{equation}
M_2=\left[ \begin{array}{c||cc||cc} R_1^\prime & I & 0  & 0 & C_1  \\
R_2^\prime & 0 & 0 & 0 & C_2 \end{array} \right]
\end{equation}
If we now apply the Hadamard gate to the last $n-k$ qubits this transforms $M_2$ into
\begin{equation}
M_3=\left[ \begin{array}{c||cc||cc} R_1^\prime & I & C_1  & 0 & 0  \\
R_2^\prime & 0 & C_2 & 0 &0  \end{array} \right]
\end{equation}
We can now repeat the Gaussian elimination on the $X$ component of this matrix, terminating if we ever obtain a row or more corresponding to $-I$.  This will produce
\begin{equation}
M_4=\left[ \begin{array}{c||cc||cc} R_3& I & 0 & 0 & 0  \\
0 & 0 & 0 & 0 &0  \end{array} \right]
\end{equation}
where $I$ is now a $k^\prime$ by $k^\prime$ identity matrix, and we have used the fact that the global phase of the lower component must be zero or else we would have terminated upon attaining a $-I$ row.  Thus we have shown that it is possible to transform an instance of the problem given by $M$ into an new equivalent instance given by $M_4$ where we terminate, answering in the negative for the problem, if we ever obtained a $-I$ row.  Finally we see that this problem is nothing more than the commuting Hamiltonian problem for a bunch of single qubit $\pm X$ operators.  This problem always has a solution to condition (a) given by an appropriate tensor product of computational basis states such that each $\pm X$ acts as trivially on this basis state.  In other words we will always answer yes if we make it all the way to the end of the reduction to $M_4$ and only answer no if somewhere along the line we obtained a $-I$ row. 

Each of the operations described above can clearly be done in polynomial time~\cite{PMH03,AG04}. This in turn completes the proof that the $k$-local Pauli Commuting Hamiltonian problem is in P.  

\section{Conclusion}
The $k$-local Hamiltonian problem is QMA-complete for $k\ge2$ (it is trivially in P when $k=1$) and it is still an open question whether the general commuting Hamiltonian is in NP (and hence NP-complete).  Here we restricted ourselves to the case that when the Hamiltonian is a sum of commuting terms which are made up of purely tensor product of Pauli matrices, and proved that the $k$-local Pauli Commuting Hamiltonian problem is actually in P.  The fundamental reason for this result is that commuting Pauli operators have eigenstates that are describable within the stabilizer formalism which itself is amenable to polynomial time classical manipulations.  This means in particular that for any $k$-local Pauli Commuting Hamiltonian problem, we could transform the individual terms into some basis where the operators are all simultaneously diagonalized, and in this basis the problem is then a simple XOR-SAT instance.  In comparison to other classes of commuting Hamiltonians that have been considered this problem ended up in P because not only did the eigenstates have an efficient description in terms of stabilizers, but this structure could be mapped onto a classic problem known to be in P (XOR-SAT).  

As we noted the Hamiltonians in the $k$-local Pauli Commuting Hamiltonian problem can exhibit ground states that have topological entanglement~\cite{Kitaev:03a}.  An important open question is whether this is true for other systems that can exhibit topological entanglement.  The $3$-local Commuting Hamiltonian system doesn't possess topological order and is in NP~\cite{AE11}. However $4$-local Commuting Hamiltonian systems do have topological order.  Here the $k$-local Pauli Commuting Hamiltonian system has topological order, but this problem was actually in P.  We therefore conclude that the complexity of general $k$-local Commuting Hamiltonian problem must depend on something more specific that whether or not a quantum system can exhibit topological order.

\begin{acknowledgments}
JY would like to thank Aram Harrow for his instructive discussion on the interpretation of the complexity result with the author.  This work was supported under NSF grants 0803478, 0829937, and 0916400 and by DARPA under QuEST grant FA-9550-09-1-0044. 
\end{acknowledgments}

\bibliography{commutingref}

\begin{thebibliography}{19}%
\makeatletter
\providecommand \@ifxundefined [1]{%
 \@ifx{#1\undefined}
}%
\providecommand \@ifnum [1]{%
 \ifnum #1\expandafter \@firstoftwo
 \else \expandafter \@secondoftwo
 \fi
}%
\providecommand \@ifx [1]{%
 \ifx #1\expandafter \@firstoftwo
 \else \expandafter \@secondoftwo
 \fi
}%
\providecommand \natexlab [1]{#1}%
\providecommand \enquote  [1]{``#1''}%
\providecommand \bibnamefont  [1]{#1}%
\providecommand \bibfnamefont [1]{#1}%
\providecommand \citenamefont [1]{#1}%
\providecommand \href@noop [0]{\@secondoftwo}%
\providecommand \href [0]{\begingroup \@sanitize@url \@href}%
\providecommand \@href[1]{\@@startlink{#1}\@@href}%
\providecommand \@@href[1]{\endgroup#1\@@endlink}%
\providecommand \@sanitize@url [0]{\catcode `\\12\catcode `\$12\catcode
  `\&12\catcode `\#12\catcode `\^12\catcode `\_12\catcode `\%12\relax}%
\providecommand \@@startlink[1]{}%
\providecommand \@@endlink[0]{}%
\providecommand \url  [0]{\begingroup\@sanitize@url \@url }%
\providecommand \@url [1]{\endgroup\@href {#1}{\urlprefix }}%
\providecommand \urlprefix  [0]{URL }%
\providecommand \Eprint [0]{\href }%
\@ifxundefined \urlstyle {%
  \providecommand \doi  [0]{\begingroup \@sanitize@url \@doi}%
  \providecommand \@doi [1]{\endgroup \@@startlink {\doibase
  #1}doi:\discretionary {}{}{}#1\@@endlink }%
}{%
  \providecommand \doi  [0]{doi:\discretionary{}{}{}\begingroup
  \urlstyle{rm}\Url }%
}%
\providecommand \doibase [0]{http://dx.doi.org/}%
\providecommand \Doi [0]{\begingroup \@sanitize@url \@Doi }%
\providecommand \@Doi  [1]{\endgroup\@@startlink{\doibase#1}\@@Doi}%
\providecommand \@@Doi [1]{#1\@@endlink}%
\providecommand \selectlanguage [0]{\@gobble}%
\providecommand \bibinfo  [0]{\@secondoftwo}%
\providecommand \bibfield  [0]{\@secondoftwo}%
\providecommand \translation [1]{[#1]}%
\providecommand \BibitemOpen [0]{}%
\providecommand \bibitemStop [0]{}%
\providecommand \bibitemNoStop [0]{.\EOS\space}%
\providecommand \EOS [0]{\spacefactor3000\relax}%
\providecommand \BibitemShut  [1]{\csname bibitem#1\endcsname}%
\bibitem [{\citenamefont {Feynman}(1982)}]{Feynman82}%
  \BibitemOpen
  \bibfield  {author} {\bibinfo {author} {\bibfnamefont {R.~P.}\ \bibnamefont
  {Feynman}},\ }\href@noop {} {\bibfield  {journal} {\bibinfo  {journal}
  {International Journal of Theoretical Physics},\ }\textbf {\bibinfo {volume}
  {21}},\ \bibinfo {pages} {467} (\bibinfo {year} {1982})}\BibitemShut
  {NoStop}%
\bibitem [{\citenamefont {Kitaev}(1999)}]{KitaevTalk99}%
  \BibitemOpen
  \bibfield  {author} {\bibinfo {author} {\bibfnamefont {A.}~\bibnamefont
  {Kitaev}},\ }\href@noop {} {\enquote {\bibinfo {title} {Lecture given at
  {H}ebrew {U}niversity, {J}erusalem, {I}srael},}\ } (\bibinfo {year}
  {1999})\BibitemShut {NoStop}%
\bibitem [{\citenamefont {Kitaev}\ \emph {et~al.}(2002)\citenamefont {Kitaev},
  \citenamefont {Shen},\ and\ \citenamefont {Vyalyi}}]{KSV02}%
  \BibitemOpen
  \bibfield  {author} {\bibinfo {author} {\bibfnamefont {A.~Y.}\ \bibnamefont
  {Kitaev}}, \bibinfo {author} {\bibfnamefont {A.~H.}\ \bibnamefont {Shen}}, \
  and\ \bibinfo {author} {\bibfnamefont {M.~N.}\ \bibnamefont {Vyalyi}},\
  }\href@noop {} {\emph {\bibinfo {title} {Classical and quantum
  computation}}},\ volume 47 of Graduate Studies in Mathematics\ (\bibinfo
  {publisher} {AMS},\ \bibinfo {address} {Providence, RI},\ \bibinfo {year}
  {2002})\BibitemShut {NoStop}%
\bibitem [{\citenamefont {Aharonov}\ and\ \citenamefont {Naveh}(2002)}]{AN02}%
  \BibitemOpen
  \bibfield  {author} {\bibinfo {author} {\bibfnamefont {D.}~\bibnamefont
  {Aharonov}}\ and\ \bibinfo {author} {\bibfnamefont {T.}~\bibnamefont
  {Naveh}},\ }\href@noop {} {\bibfield  {journal} {\bibinfo  {journal}
  {arXiv:quant-ph/0210077}} (\bibinfo {year} {2002})}\BibitemShut {NoStop}%
\bibitem [{\citenamefont {Bravyi}\ and\ \citenamefont {Vyalyi}(2005)}]{BV04}%
  \BibitemOpen
  \bibfield  {author} {\bibinfo {author} {\bibfnamefont {S.}~\bibnamefont
  {Bravyi}}\ and\ \bibinfo {author} {\bibfnamefont {M.}~\bibnamefont
  {Vyalyi}},\ }\href@noop {} {\bibfield  {journal} {\bibinfo  {journal}
  {Quantum Inf. and Comp.},\ }\textbf {\bibinfo {volume} {5}},\ \bibinfo
  {pages} {187} (\bibinfo {year} {2005})}\BibitemShut {NoStop}%
\bibitem [{\citenamefont {Bravyi}(2006)}]{Bravyi06}%
  \BibitemOpen
  \bibfield  {author} {\bibinfo {author} {\bibfnamefont {S.}~\bibnamefont
  {Bravyi}},\ }\href@noop {} {\bibfield  {journal} {\bibinfo  {journal}
  {arXiv:quant-ph/0602108v1}} (\bibinfo {year} {2006})}\BibitemShut {NoStop}%
\bibitem [{\citenamefont {Kempe}\ \emph {et~al.}(2006)\citenamefont {Kempe},
  \citenamefont {Kitaev},\ and\ \citenamefont {Regev}}]{KKR06}%
  \BibitemOpen
  \bibfield  {author} {\bibinfo {author} {\bibfnamefont {J.}~\bibnamefont
  {Kempe}}, \bibinfo {author} {\bibfnamefont {A.}~\bibnamefont {Kitaev}}, \
  and\ \bibinfo {author} {\bibfnamefont {O.}~\bibnamefont {Regev}},\
  }\href@noop {} {\bibfield  {journal} {\bibinfo  {journal} {SIAM J. Comput.},\
  }\textbf {\bibinfo {volume} {35}},\ \bibinfo {pages} {1070} (\bibinfo {year}
  {2006})}\BibitemShut {NoStop}%
\bibitem [{\citenamefont {Aharonov}\ \emph {et~al.}(2007)\citenamefont
  {Aharonov}, \citenamefont {van Dam}, \citenamefont {Kempe}, \citenamefont
  {Landau}, \citenamefont {Lloyd},\ and\ \citenamefont {Regev}}]{ADKLLR07}%
  \BibitemOpen
  \bibfield  {author} {\bibinfo {author} {\bibfnamefont {D.}~\bibnamefont
  {Aharonov}}, \bibinfo {author} {\bibfnamefont {W.}~\bibnamefont {van Dam}},
  \bibinfo {author} {\bibfnamefont {J.}~\bibnamefont {Kempe}}, \bibinfo
  {author} {\bibfnamefont {Z.}~\bibnamefont {Landau}}, \bibinfo {author}
  {\bibfnamefont {S.}~\bibnamefont {Lloyd}}, \ and\ \bibinfo {author}
  {\bibfnamefont {O.}~\bibnamefont {Regev}},\ }\href@noop {} {\bibfield
  {journal} {\bibinfo  {journal} {SIAM J. Comput.},\ }\textbf {\bibinfo
  {volume} {37}},\ \bibinfo {pages} {166} (\bibinfo {year} {2007})}\BibitemShut
  {NoStop}%
\bibitem [{\citenamefont {Oliveira}\ and\ \citenamefont
  {Terhal}(2009)}]{Oliveira09}%
  \BibitemOpen
  \bibfield  {author} {\bibinfo {author} {\bibfnamefont {R.}~\bibnamefont
  {Oliveira}}\ and\ \bibinfo {author} {\bibfnamefont {B.~M.}\ \bibnamefont
  {Terhal}},\ }\href@noop {} {\bibfield  {journal} {\bibinfo  {journal} {Quant.
  Inf. Comput.},\ }\textbf {\bibinfo {volume} {8}},\ \bibinfo {pages} {900}
  (\bibinfo {year} {2009})}\BibitemShut {NoStop}%
\bibitem [{\citenamefont {Schuch}(2011)}]{Schuch11}%
  \BibitemOpen
  \bibfield  {author} {\bibinfo {author} {\bibfnamefont {N.}~\bibnamefont
  {Schuch}},\ }\href@noop {} {\bibfield  {journal} {\bibinfo  {journal}
  {arXiv:1105.2843v1}} (\bibinfo {year} {2011})}\BibitemShut {NoStop}%
\bibitem [{\citenamefont {Aharonov}\ and\ \citenamefont {Eldar}(2011)}]{AE11}%
  \BibitemOpen
  \bibfield  {author} {\bibinfo {author} {\bibfnamefont {D.}~\bibnamefont
  {Aharonov}}\ and\ \bibinfo {author} {\bibfnamefont {L.}~\bibnamefont
  {Eldar}},\ }\href@noop {} {\bibfield  {journal} {\bibinfo  {journal}
  {arXiv:1102.0770v2}} (\bibinfo {year} {2011})}\BibitemShut {NoStop}%
\bibitem [{\citenamefont {Kitaev}\ and\ \citenamefont
  {Preskill}(2006)}]{Kitaev:06a}%
  \BibitemOpen
  \bibfield  {author} {\bibinfo {author} {\bibfnamefont {A.}~\bibnamefont
  {Kitaev}}\ and\ \bibinfo {author} {\bibfnamefont {J.}~\bibnamefont
  {Preskill}},\ }\href@noop {} {\bibfield  {journal} {\bibinfo  {journal}
  {Phys. Rev. Lett.},\ }\textbf {\bibinfo {volume} {96}},\ \bibinfo {pages}
  {110404} (\bibinfo {year} {2006})}\BibitemShut {NoStop}%
\bibitem [{\citenamefont {Gottesman}(1997)}]{Gottesman97}%
  \BibitemOpen
  \bibfield  {author} {\bibinfo {author} {\bibfnamefont {D.}~\bibnamefont
  {Gottesman}},\ }\href@noop {} {\bibfield  {journal} {\bibinfo  {journal}
  {Ph.D. Thesis, arXiv:quant-ph/9705052v1}} (\bibinfo {year}
  {1997})}\BibitemShut {NoStop}%
\bibitem [{\citenamefont {Kitaev}(2003)}]{Kitaev:03a}%
  \BibitemOpen
  \bibfield  {author} {\bibinfo {author} {\bibfnamefont {A.}~\bibnamefont
  {Kitaev}},\ }\href@noop {} {\bibfield  {journal} {\bibinfo  {journal} {Ann.
  of Phys.},\ }\textbf {\bibinfo {volume} {303}},\ \bibinfo {pages} {2}
  (\bibinfo {year} {2003})}\BibitemShut {NoStop}%
\bibitem [{\citenamefont {Aharonov}\ \emph {et~al.}(2009)\citenamefont
  {Aharonov}, \citenamefont {Gottesman}, \citenamefont {Irani},\ and\
  \citenamefont {Kempe}}]{AGIK07}%
  \BibitemOpen
  \bibfield  {author} {\bibinfo {author} {\bibfnamefont {D.}~\bibnamefont
  {Aharonov}}, \bibinfo {author} {\bibfnamefont {D.}~\bibnamefont {Gottesman}},
  \bibinfo {author} {\bibfnamefont {S.}~\bibnamefont {Irani}}, \ and\ \bibinfo
  {author} {\bibfnamefont {J.}~\bibnamefont {Kempe}},\ }\href@noop {}
  {\bibfield  {journal} {\bibinfo  {journal} {Comm. Math. Physics},\ }\textbf
  {\bibinfo {volume} {287}},\ \bibinfo {pages} {41} (\bibinfo {year}
  {2009})}\BibitemShut {NoStop}%
\bibitem [{\citenamefont {Barahona}(1982)}]{Barahona:82a}%
  \BibitemOpen
  \bibfield  {author} {\bibinfo {author} {\bibfnamefont {F.}~\bibnamefont
  {Barahona}},\ }\href@noop {} {\bibfield  {journal} {\bibinfo  {journal} {J.
  Phys. A: Math. Gen.},\ }\textbf {\bibinfo {volume} {15}},\ \bibinfo {pages}
  {3241} (\bibinfo {year} {1982})}\BibitemShut {NoStop}%
\bibitem [{\citenamefont {Gottesman}(1999)}]{Gottesman99}%
  \BibitemOpen
  \bibfield  {author} {\bibinfo {author} {\bibfnamefont {D.}~\bibnamefont
  {Gottesman}},\ }in\ \href@noop {} {\emph {\bibinfo {booktitle} {Proceedings
  of the XXII International Colloquium on Group Theoretical Methods in
  Physics}}},\ \bibinfo {editor} {edited by\ \bibinfo {editor} {\bibfnamefont
  {R.~D.}\ \bibnamefont {S.~P.~Corney}}\ and\ \bibinfo {editor} {\bibfnamefont
  {P.~D.}\ \bibnamefont {Jarvis}}}\ (\bibinfo  {publisher} {International
  Press},\ \bibinfo {address} {Cambridge, MA},\ \bibinfo {year} {1999})\ pp.\
  \bibinfo {pages} {32--43}\BibitemShut {NoStop}%
\bibitem [{\citenamefont {Aaronson}\ and\ \citenamefont
  {Gottesman}(2004)}]{AG04}%
  \BibitemOpen
  \bibfield  {author} {\bibinfo {author} {\bibfnamefont {S.}~\bibnamefont
  {Aaronson}}\ and\ \bibinfo {author} {\bibfnamefont {D.}~\bibnamefont
  {Gottesman}},\ }\href@noop {} {\bibfield  {journal} {\bibinfo  {journal}
  {Phys. Rev. A},\ }\textbf {\bibinfo {volume} {70}},\ \bibinfo {pages}
  {052328} (\bibinfo {year} {2004})}\BibitemShut {NoStop}%
\bibitem [{\citenamefont {Patel}\ \emph {et~al.}(2003)\citenamefont {Patel},
  \citenamefont {Markov},\ and\ \citenamefont {Hayes}}]{PMH03}%
  \BibitemOpen
  \bibfield  {author} {\bibinfo {author} {\bibfnamefont {K.}~\bibnamefont
  {Patel}}, \bibinfo {author} {\bibfnamefont {I.}~\bibnamefont {Markov}}, \
  and\ \bibinfo {author} {\bibfnamefont {J.}~\bibnamefont {Hayes}},\
  }\href@noop {} {\bibfield  {journal} {\bibinfo  {journal}
  {arXiv:quant-ph/0302002v1}} (\bibinfo {year} {2003})}\BibitemShut {NoStop}%
\end{thebibliography}%

\end{document}